\documentclass[aps,pre,showpacs,amsmath,amsfonts,amssymb,superscriptaddress,preprint]{revtex4-1}

\pdfoutput=1

\usepackage[english]{babel}
\usepackage[pdftex]{graphicx}
\usepackage{tikz}
\usepackage{rotating}
\usepackage[pdftex,hypertexnames=true,linktocpage=true]{hyperref} 
\hypersetup{colorlinks=true,linkcolor=blue,anchorcolor=blue,citecolor=magenta,filecolor=blue,urlcolor=blue,bookmarksnumbered=false,pdfview=FitB}

\usepackage{amsthm}
\usepackage{framed}
\usepackage{amssymb}
\usepackage{amsmath,mathrsfs}
\usepackage{hhline}
\usepackage{ulem}
\usepackage{fancyhdr}
\usepackage{simplewick}

\everymath{\displaystyle}

\newcommand{\beq}{\begin{equation}}
\newcommand{\eeq}{\end{equation}}

\newcommand{\bi}{\begin{itemize}}
\newcommand{\ei}{\end{itemize}}

\newcommand{\affA}{Aix-Marseille Universit\'e, Marseille, France}
\newcommand{\affB}{CNRS Centre de Physique Th\'eorique UMR7332,
13288 Marseille, France}
\newcommand{\affC}{Centre d'Immunologie de Marseille-Luminy, Aix Marseille Universit\'e  UM2,
Inserm, U1104, CNRS UMR7280, 13288 Marseille, France}

\begin{document}

\title{Random walk of passive tracers among randomly moving obstacles}

\author{Matteo Gori}
\email{gori6matteo@gmail.com ;  corresponding author }
\affiliation{\affA}\affiliation{\affB}
\author{Irene Donato}
\email{irene.irened@gmail.com}
\affiliation{\affA}\affiliation{\affB}
\author{Elena Floriani}
\email{elena.floriani@cpt.univ-mrs.fr}
\affiliation{\affA}\affiliation{\affB}
\author{Ilaria Nardecchia}
\email{i.nardecchia@gmail.com}
\affiliation{\affB}\affiliation{\affC}
\author{Marco Pettini}
\email{pettini@cpt.univ-mrs.fr}
\affiliation{\affA}\affiliation{\affB}

\begin{abstract}
Background: This study is mainly motivated by the need of understanding how the diffusion behaviour of a biomolecule (or even of a larger object) is affected by other moving macromolecules, organelles, and so on, inside a living cell, whence the possibility of understanding whether or not a randomly walking biomolecule is also subject to a long-range force field driving it to its target.

Method: By means of the Continuous Time Random Walk (CTRW) technique the topic of random walk in random environment is here considered in the case of a passively diffusing particle in a crowded environment made of randomly moving and interacting obstacles. 

Results: The relevant physical quantity  which is worked out is the diffusion coefficient of the passive tracer which is computed as a function of the average inter-obstacles distance. 

Coclusions: The results reported here suggest that if a biomolecule, let us call it a test molecule, moves towards its target in the presence of other independently interacting molecules,  its motion can be considerably slowed down.  Hence, if such a slowing down could compromise the efficiency of the task to be performed by the test molecule, some accelerating factor would be required. Intermolecular electrodynamic forces are good candidates as accelerating factors because they can act  at a long distance in a medium like the cytosol despite its ionic strength.
\end{abstract}

\date{\today}


\maketitle

\section{Introduction}
The topic of random walk in random environment (RWRE) has been the object of extensive studies  during the last four decades and is of great interest to mathematics, physics and several applications. There is a huge literature on numerical, theoretical, and rigorous analytical results. The subject has been pioneered both through applications, as is the case of the models introduced to describe DNA replication \cite{chernov}, or through more abstract models in the field of probability theory \cite{harris}. One can find in Ref.\cite{solomon} the definition of the mathematical framework of RWRE and since then a vast body of results has been built for both static and dynamic random environments, to mention just a few of them see \cite{boldri1,boldri2,liver1,liver2} and the references therein quoted.

In a biophysical context this kind of  problems is referred to as "macromolecular crowding" which, among other issues,  encompasses the effects of  excluded volume on molecular diffusion and biochemical reaction rates within living cells. Another kind of biophysical application of RWRE is related with single-particle tracking experiments allowing to measure the diffusion coefficient of an individual particle (protein or lipid) on the cell surface. The knowledge of single-trajectory diffusion coefficients is useful as a measure of the heterogeneity of the cell membrane and requires to model hindered diffusion conditions \cite{saxton}. 

To give another example among a huge number of processes in living matter, during B lymphocyte development, immunoglobulin heavy-chain variable, diversity, and joining segments assemble to generate a diverse antigen receptor repertoire. Spatial confinement related with diffusion hindrance from the surrounding network of proteins and chromatin fibers is the dominant parameter that determines the frequency of encounters of the above mentioned segments. When these particles encounter obstacles present at high concentration, the particles motions become subdiffusive \cite{cell} as described by the continuous time random walk (CTRW) model \cite{saxton,montroll}. 

Within the "macromolecular crowding" framework, the typical problem tackled to study biochemical reaction kinetics is that of finding how the diffusion of certain interacting particles is affected by the presence of Brownian  non interacting particles (crowding agents) of a different kind.

In the present paper we consider, so to speak, the "dual" situation, that is,  the diffusion of  Brownian passive tracers in presence of other species of interacting particles now playing the role of crowding agents. The reason for considering this problem stems from the need of estimating how the encounter time of a given macromolecule (passive tracer) with its cognate partner, say a transcription factor diffusing towards is target on the DNA, is affected by the surrounding particles intervening in other biochemical reactions. As we shall  see  in the following, a molecule diffusing through a medium crowded by  moving obstacles (molecules) interacting among each other can be considerably slowed down. If, say, a passive tracer (molecule) had to reach its target in a short time by diffusing through a crowded medium that prevents it from performing its task, then the help of some long-distance force field that attracts the tracer towards the target would be necessary.

In other words, the outcomes of the present work could be useful tools to infer whether intermolecular electrodynamic forces acting at a long distance can be at work in living matter. First steps in this perspective are being done in some recent works \cite{pre1,pre2,pre3,fels}.

\section{Methods: Continuous Time Random Walk formalism} \label{ctrw}
One of the many ways of modelling diffusive behaviour is by Continuous Time Random Walk (CTRW) \cite{ZK93,KBS87}. This framework is mainly used to extend the description of Brownian motion to anomalous transport, in order to deal with subdiffusive or superdiffusive behaviour in connection with L\'evy processes,  but it can of course be used to describe the simpler and more frequent case of normal diffusion. In this paper, we focus on cases where diffusion of tracers and interacting molecules is indeed Gaussian, so that a diffusion coefficient can be defined. 

Let us consider a population of independent particles $A$, and let us suppose that their motion can be modelled as a sequence of motion events that take place in three dimensions and in continuous time.  The extension to two and one dimensions is trivial, and in the literature (see for exemple  \cite{ZK93}) calculations are often carried out in one dimension. 

In the CTRW framework the random walk is specified by $\psi({\bf r}, t)$, the probability density of making a displacement ${\bf r}$ in time $t$ in a single motion event. The normalization condition on $\psi({\bf r}, t)$ is
\beq
\int_0^{+\infty} dt \int d^3{\bf r} \;\psi({\bf r}, t) = 1 
\eeq
In many applications of CTRW $\psi({\bf r}, t)$ is decoupled so that there is no correlation between the displacement ${\bf r}$ and the time interval $t$:
\begin{equation} \label{lambdapsi}
\psi({\bf r}, t) = \Lambda({\bf r})\,\psi(t) 
\end{equation}

Here we rather consider the formulation where space and time are coupled, thus expressing the fact that the particles move with a given velocity during single motion events; this amounts to introducing a conditional probability $p({\bf r}| t) $, i.e., the probability that a given displacement ${\bf r}$ takes place in a time $t$
\beq \label{psirt}
\psi({\bf r}, t) = \Lambda({\bf r})\; p({\bf r}| t) = \Lambda({\bf r})\; \delta\left(t - \frac{\left|{\bf r}\right|}{\left|{\bf v}({\bf r})\right|} \right) 
\eeq
Normalization requires that
\begin{equation}
\int d^3{\bf r} \;\Lambda({\bf r}) = 1 
\end{equation}
We take the velocity to be constant in magnitude
\beq
\psi({\bf r}, t) = \Lambda({\bf r})\; \delta\left(t - \frac{\left|{\bf r}\right|}{v_0}\right) 
\eeq
Furthermore, we consider isotropic systems, which implies that the distribution $\Lambda({\bf r})$ is a function of $r=|{\bf r}|$ only. We write it in the form 
\begin{equation}
\Lambda({\bf r}) = \dfrac{\lambda(r)}{4\pi r^2}
\end{equation}
with the normalization condition
\begin{equation}
\int_0^{+\infty} dr\;\lambda(r) = 1
\end{equation}
which allows to rewrite $\psi({\bf r}, t)$ as
\begin{equation} \label{psiphi}
\psi({\bf r},t) = \frac{\lambda(r)}{4\pi\,r^2}\,\delta\left(t - \frac{\left|{\bf r}\right|}{ v_0}\right) = \frac{\phi(t)}{4\pi\,(v_0 t)^2}\,\delta\left(\left|{\bf r}\right| - v_0 t\right)
\end{equation}
where  $\phi(t)$ is the free-flight or waiting time distribution which represents the probability density function for a random walker to keep the same direction of its velocity during a time $t$ and is the fundamental quantity for the description of our isotropic system. The free-flight distribution satisfies the relations
\begin{equation} \label{philambda}
\phi(t) = v_0 \lambda(v_0 t)\qquad\int_0^{+\infty} dt \,\,\phi(t) = 1
\end{equation}

Starting from these quantities, one can compute the Fourier-Laplace transform of the probability density $P({\bf r}, t)$ for a particle to be at the position ${\bf r}$ at time $t$, and consequently calculate the diffusion properties.  This is done in the Appendix, where we generalise to three dimensions the analysis carried out in \cite{ZK93} for the one-dimensional case, considering two slightly different versions of the CTRW: 

\begin{itemize}
\item[(i)]
The Velocity Model, in which each particle $A$ moves with constant velocity $v_0$ between two turning points; at a turning point, a new direction and a new length of flight are taken according to the probability density $\Lambda({\bf r})$.
\item[(ii)]
The Jump Model, in which each particle waits at a particular location before instantaneously moving to the next one, the displacement being chosen according to the probability density $\Lambda({\bf r})$, the waiting time for a jump to take place being $|{\bf r}|/v_0$.
\end{itemize}

The expression of $P({\bf r}, t)$ is formally different for these two versions of the CTRW, but from their definition it appears that the two models are equivalent in the long time limit.    

As a general remark on other possible applications of our work, this CTRW description where space ans time are coupled (see equation (\ref{psirt})) allows to model situations of Gaussian diffusion but also of enhanced diffusion (where $\left\langle r^2(t)\right\rangle\simeq t^{\alpha}$ with $\alpha>1$) \cite{KBS87}, because it can describe cases where the particles keep the same velocity for very long times (if the free-flight distribution $\phi(t)$ decays slowly, typically as an inverse power law).  

We get normal diffusion as soon as $\phi(t)$ has a finite second moment. In this case, the long time behaviour of the mean square displacement, and hence of the diffusion coefficient, is, both for the Velocity and Jump models (see the Appendix)

\beq \label{r2t}
\left\langle r^2(t)\right\rangle = \int d^3{\bf r}\; |{\bf r}|^2 P({\bf r}, t) \simeq \frac{v_0^2\langle t^2\rangle_{\phi}}{\langle t\rangle_{\phi}}\;t
\eeq
where $\langle t^n\rangle_{\phi}$ is the $n$-th moment of the distribution $\phi(t)$ defined by Equations (\ref{psiphi}) and (\ref{philambda}):
\beq
\langle t^n\rangle = \int_0^{+\infty} dt\,t^n\phi(t) 
\eeq

The diffusion coefficient is then given by

\beq \label{D}
D = \lim_{t\rightarrow\infty} \frac{\langle r^2(t)\rangle}{6t} =\frac{v_0^2\langle t^2\rangle_{\phi}}{6\langle t\rangle_{\phi}}
\eeq

Let us notice that the same CTRW formalism can also describe subdiffusion (where $\left\langle r^2(t)\right\rangle\simeq t^{\alpha}$ with $\alpha<1$) \cite{KBS87}. This can be obtained by considering a version of the Jump Model where space and time are decoupled, as in equation (\ref{lambdapsi}): particles remain at a particular location for times distributed according to $\psi(t)$ and make instantaneous jumps on distances distributed according to $\Lambda({\bf r})$. Subdiffusion is obtained as soon as $\Lambda({\bf r})$ has finite second moment while the first moment of the waiting time distribution $\psi(t)$ diverges.

\section{Diffusion of independent tracers in the presence of interacting obstacles}

If we adopt the CTRW description of diffusion of the preceding section, then the main quantity to consider is $\phi_A(t)$, the probability density function that a random walker $A$ keeps the same direction of velocity during a time $t$. 

We will refer to "unperturbed" diffusion if $A$ is the only species present in a solution, and we will denote the free-flight time distribution of the unperturbed case by $\phi_{0_A}(t)$.
The discussion in the preceding section gives 
\beq \label{D0_ctrw}
D_{0_A}= \frac{v_{0_A}^2\langle t^2\rangle_{\phi_{0_A}}}{6\langle t\rangle_{\phi_{0_A}}}\;
\eeq
which is also independently given by Einstein's relation 
\beq \label{D0_einstein}
D_{0_A} = \frac{kT}{\gamma_A}
\eeq
where $k$ is the Boltzmann constant, $T$ is the temperature
and $\gamma_A$ is the friction coefficient for A-particles according to Stokes' Law:
\begin{equation}
\label{eq:StokesLaw}
\gamma =6\pi\,R_A\,\eta
\end{equation}
where $R_A$ is the hydrodynamic radius of the diffusing particles and $\eta$ is the viscosity of the medium where the particles diffuse. 

Moreover, we can estimate the typical particle velocity using equipartition of energy
\beq \label{v02_equipart}
v_{0_A}^2 = \frac{3kT}{m_A}
\eeq
where $m_A$ is the mass of a particle $A$.

So, if we interpret $\phi_{0_A}(t)$ as the free-flight time distribution between Brownian collisions of the particles $A$ on the molecules of the medium, then equations (\ref{D0_ctrw}), (\ref{D0_einstein}), (\ref{v02_equipart}) imply that the two first moments of $\phi_{0_A}$ must satisfy the relation
\beq \label{t2t}
\frac{\langle t^2\rangle_{\phi_{0_A}}}{2\langle t\rangle_{\phi_{0_A}}} = \frac{m_A}{\gamma_A}
\eeq

As stated in the Introduction, the physical situation we are interested in is that one where another population of particles, say $B$-particles, is also present in the solution. Particles $B$ are supposed to diffuse and mutually interact, but there is no interaction at a distance between them and the particles $A$. It is reasonable to suppose that the diffusive and dynamic properties of these moving obstacles $B$ induce changes in the diffusive properties of the $A$-particles which can be thus seen as passive tracers.

We want to model how the $B$-particles affect the diffusion properties of the $A$-particles by resorting to a suitable modification of the CTRW probability distribution $\phi_A(t)$. The amount of the modification will of course depend on the concentration $C_B$ (or equivalently on the average distance $d=C_B^{-1/3}$) of obstacles. Our goal is to estimate with simple arguments the dependence on the average distance $d$ between any pair of obstacles of the ratio $D_A/D_{0_A}$ between perturbed and unperturbed diffusion coefficients.

We always assume that
\begin{equation}
C_A \ll C_B
\end{equation}
so that the A-particles can be regarded as tracers: any A-particle does not influence the dynamics of the obstacles and of the other tracers.

\subsection{Modification of the microscopic free-flight time distribution} \label{micro}

If the concentration $C_B$ of the obstacles $B$ is low enough, in the sense that their average distance $d$ is such that
\begin{equation}
d\gg \sqrt{\int_0^{+\infty} dr \; r^2 \lambda_0(r)}
\end{equation}
we can consider that the diffusion of $A$-particles is not perturbed by the presence of the obstacles $B$; thus for the waiting time distribution we will have $\phi_A(t)\simeq \phi_{0_A}(t)$, and, consequently, $D_A\simeq D_{0_A}$.

As the concentration of $B$-particles grows, the diffusion of $A$-particles is affected accordingly, and this is described by a modification of $\phi_A(t)$. It is reasonable to suppose that $\phi_A(t)$ will be close to $\phi_{0_A}(t)$ at sufficiently short times, i.e., for displacements small enough that a tracer $A$ does not "see" any obstacle $B$, and that $\phi_A(t)$ will be reduced with respect to the unperturbed $\phi_{0_A}(t)$ at long times, because long free displacements are likely to be interrupted by the presence of obstacles. 

Following this idea, we model the waiting time distribution as follows: we call $T_d$ the characteristic time of flight at which a tracer $A$ begins to "see" the obstacles $B$, where $T_d$ will depend of course on the typical distance $d\simeq C_B^{-1/3}$ between the $B$-particles. We then make the simplest assumption that $\phi_A(t)$ coincides (except for a normalisation factor) with $\phi_{0_A}(t)$ for times smaller than $T_d$ and is is zero for times larger than $T_d$.


We take the unperturbed distribution $\phi_{0_A}(t)$ to be exponentially decreasing
\beq \label{phi0}
\phi_{0_A}(t) = \dfrac{1}{\tau_A}\,e^{-t/\tau_A}
\eeq
where, using equation \eqref{t2t}:
\begin{equation}
\label{eq:taufree}
\tau_A=\dfrac{m_A}{\gamma_A}
\end{equation}

So, we write the modified probability density $\phi_A(t)$ for passive tracers (A-particles) in presence of interacting moving obstacles (B-particles) as: 
\beq \label{phi}
\phi_A(t) = \dfrac{e^{-t/\tau_A}}{\tau_A(1-e^{-T_d/\tau_A})}\;\;\;\;\;\mbox{if}\;\; t<T_d\;,\;\;\;\;\;\;\;\;\;\; 
\phi_A(t)=0\;\;\;\;\;\mbox{if}\;\; t \ge T_d
\eeq
If we compute the diffusion coefficient using equation (\ref{D}) and expressions \eqref{phi}, \eqref{phi0} we get
\beq \label{DD0}
\dfrac{D_A}{D_{0_A}} = \dfrac{\langle t^2\rangle_{\phi_A}}{\langle t\rangle_{\phi_A}}\cdot\dfrac{\langle t\rangle_{\phi_{0_A}}}{\langle t^2\rangle_{\phi_{0_A}}}
= 1 - \dfrac{x^2}{2\,(e^x -1 -x)}\;\;\;\;\;\mbox{where}\;\;x=x(d)=\dfrac{T_d}{\tau_A}
\eeq
which is a function of the ratio between the transition time $T_d$ and the characteristic timescale $\tau_A$ of the non perturbed waiting time distribution.

The issue is now to establish the dependence of the transition time $T_d$ (and consequently, of the parameter $x$) on the average distance $d$ between obstacles.

We assume that the $B$-molecules (obstacles) diffuse: we could apply to them the same CTRW description with velocity $v_{0_B}$ and waiting time distribution $\phi_B(t)$ in the case of non interacting obstacles. Moreover if the the obstacles mutually interact they have a systematic drift velocity that is due to deterministic forces acting between them. This drift velocity  depends on their mutual distance $d$, and we will call it $V_d$. If we suppose that the dynamics of the $B$-molecules  is  over-damped, a crude estimation of $V_d$ is given by $V_d \simeq F(d)/\gamma_B$, where $\gamma_B= 6\pi\,R_B\eta$ is the friction coefficient of the $B$-molecules and $F(d)$ is the module of the deterministic force between two molecules of type $B$ at a distance $d=C_B^{-1/3}$.

The transition time $T_d$ can be  roughly  estimated by considering that, if the diffusive displacement of a tracer $A$ is interrupted by the presence of the $B$-molecules, this is due to a molecule $B$ which is moving in the direction of the tracer $A$, so that
\beq \label{eqTd}
T_d \simeq \dfrac{d}{v_{0_A} + v_{0_B} + V_d} \simeq \dfrac{d}{v_{0_A} + v_{0_B} + F(d)/\gamma_B} 
\eeq
For the parameter $x$ appearing in (\ref{DD0}) this gives
\begin{equation} \label{Tdtau}
x=x(d)=\dfrac{T_d}{\tau_A}=\dfrac{d}{\dfrac{m_A}{\gamma_A} 
\left[ \sqrt{\dfrac{3kT}{m_A}} + \sqrt{\dfrac{3kT}{m_B}} + \dfrac{F(d)}{\gamma_B}\right]}
\end{equation}
where we have used equations (\ref{v02_equipart}), (\ref{t2t}) and (\ref{phi0}).

Now, some remarks are in order. The most delicate point in the procedure mentioned above to compute $D_A/D_{0_A}$ of the tracers consists in the choice of the functional form of $T_d=T_d[U(r)](d)$. Equation (\ref{eqTd}) is a rough estimate of this characteristic time because it excludes, for instance, effects due to the dimensionality of physical space where diffusion takes place (1D, 2D, etc.), the sign of interaction energy among obstacles, spatial correlation among obstacles and the possibility of multiple collisions among the molecules.

The last point entails the exclusion - from the range of validity of our model - of all the cases where $d \lesssim \min \{R_A,R_B\}$  (as in the case of densely crowded systems). For this reason we do not take into account the sizes of both tracers and obstacles at a distance $d$ from the colliding particle.

Moreover this modelization is meaningful if the transition time $T_d$ is of the same order of magnitude of the characteristic timescale $\tau_A$ of $\phi_A(t)$. Such condition is equivalent to require that the viscosity $\eta$ of the medium and the interparticle distance $d$ are sufficiently small and, possibly, the interaction strength among the obstacles is sufficiently large. To the contrary, if the parameters of the system are such that the typical time $T_d$ at which the tracers "see" the obstacles is many orders of magnitude larger than the typical time $\tau_A$ between Brownian collisions, the free-flight time distribution $\phi_{0_A}(t)$ will not be modified by the presence of the obstacles, and Equation (\ref{DD0}) will always give $D_A\simeq D_{0_A}$, as $x(d)\gg 1$ for all the accessible values of $d$. More precisely, if we look at equation (\ref{Tdtau}) for the ratio between $T_d$ and $\tau$, we see that it is reasonable to think that the presence of the $B$-particles modifies the microscopic free-flight time distribution between Brownian collisions if the product $(\gamma_A d)$ is not much larger than $\sqrt{m_A k T}$.  Unfortunately this is not true in many applications. Consider, for instance, the case of two molecular species diffusing in water ($\eta = 5.1 \times 10^{8}$ KDa $\mu$m$^{-1}$ $\mu$s$^{-1}$) at room temperature $T=300 K$, where the $A$-particles are non interacting small molecules (say a small peptide complex), and the $B$-particles  represent mutually interacting biomolecules with $m_B\simeq 20$ KDa and $R_B\simeq 2 \times 10^{-3}\,\mu$m, so that $R_A\simeq 0.5\, R_B$ and $m_A\simeq 0.025 m_B\simeq 0.5 \,\text{KDa}$. Using equations \eqref{eq:taufree} and the previous choice of physical parameters for A-particles, we obtain that $\tau_A\simeq 5\times 10^{-8}\mu$s. Moreover suppose that the $B$-particles are characterized by a net electric charge $Z_B\simeq 10$, that their mutual average distance is $d=0.05 \mu$m$\simeq 50\,R_B$, and that they interact through a  non screened electrostatic potential. This models the case of an ideal watery solution of $A$- and $B$-type particles with no Debye screening, and with $\varepsilon_{rel}\simeq 80$ (the value of the static dielectric constant of water). Using equation \eqref{v02_equipart} we see that the contribution due to thermal noise of $A$-type molecules is larger than that of the $B$-type molecules, in fact $v_{0_A} \simeq 1.2\times10^{2}\mu\text{m}\mu\text{s}^{-1} \simeq 6\, v_{0_B}$; moreover, the interaction term is negligible with respect to the velocities, as
\begin{equation}
\dfrac{F(d)}{\gamma_B}=\dfrac{Z_B^2 q^2}{\varepsilon_{rel} d^2}\dfrac{1}{\gamma_B}\simeq 7 \times10^{-3}\mu m \mu s^{-1} \simeq 3\times 10^{-3} v_{0_B}
\end{equation}
where $q$ is the elementary charge expressed in Gaussian units. Using \eqref{eqTd}, the transition time is $T_d\simeq 3\cdot 10^{-4} \mu$s, whence we get $x(d)\simeq 6 \cdot 10^{3}$.

\subsection{Modification of the rescaled free-flight time distribution} \label{rescaled}

In order to describe physical systems for which $T_d\gg \tau$ for all the accessible values of the intermolecular distance $d$, as the one described by the preceding example, we have to modify the CTRW modelization. 

Let us still model the unperturbed diffusion of tracers as a sequence of linear motion events described in the CTRW formalism by a rescaled function $\tilde{\psi}_{0_A}({\bf r},t)$ given by
\beq
\label{eqpsiA0_resc}
\tilde{\psi}_{0_A} ({\bf r},t) = \dfrac{1}{4\pi\,(\tilde{v}_{0_A} t)^2}\,\tilde{\phi}_{0_A}(t)\,\delta(\left|{\bf r}\right| - \tilde{v}_{0_A} t)
= \dfrac{1}{4\pi\,(\tilde{v}_{0_A} t)^2}\,\dfrac{e^{-t/\tilde{\tau}_A}}{\tilde{\tau}_A}\,\delta(\left|{\bf r}\right| - \tilde{v}_{0_A} t)
\eeq
where $\tilde{v}_{0_A}=\alpha_A v_{0_A}$ is a rescaled velocity and $\tilde{\tau}_A=\beta_A\tau_A$ is a rescaled characteristic timescale for diffusive motion events. The parameters $v_{0_A}$ and $\tau_{A}$ are the same as in the previous Section. If there are no interactions among obstacles (B-particles), a relation equivalent to equation \eqref{eqpsiA0_resc} can be written for each B particle
\begin{equation}
\label{eq:psiB0_resc}
\tilde{\psi}_{0_B} ({\bf r},t) = \dfrac{1}{4\pi\,(\tilde{v}_{0_B} t)^2}\,\tilde{\phi}_{0_B}(t)\,\delta(\left|{\bf r}\right| - \tilde{v}_{0_B} t)
= \dfrac{1}{4\pi\,(\tilde{v}_{0_B} t)^2}\,\dfrac{e^{-t/\tilde{\tau}_B}}{\tilde{\tau}_B}\,\delta(\left|{\bf r}\right| - \tilde{v}_{B_0} t)
\end{equation}
where, analogously to the previous case, $\tilde{v}_{0_B}=\alpha_B v_{0_B}$ and $\tilde{\tau}_B=\beta_B\tau_B$.

Of course this does not model the microscopic level, in the sense that the single motion events - whose probability is specified by $\tilde{\psi}_A ({\bf r},t)$ - are no longer the microscopic displacements between successive Brownian collisions. Rather, we focus on the motion on longer timescales $\tilde{\tau}_A$ ($\beta_A>1$) and model the diffusion of tracers as a sequence of displacements on typical distances $\tilde{v}_{A_0}\tilde{\tau}_{A}$.

The conditions on the rescaling parameters $(\alpha_A, \beta_A, \alpha_B,\beta_B)$ are then
\bi
\item the typical motion event for tracer (A-particles) takes
place between two consecutive encounters with an obstacle (B-particles); this means that the spatial scale of a typical motion event for tracers described by $\tilde{\psi}_{0_A}({\bf r},t)$ is $d$, the average distance between any two obstacles. This condition guarantees that $\tau_A$, and consequently
 $\tilde{\psi}_{0_A}({\bf r},t)$, is modified in the presence of obstacles
\beq
\label{eq:tauA_cond}
\left(\tilde{v}_{0_A}+\tilde{v}_{0_B}\right)\tilde{\tau}_A = \left(\alpha_A v_{0_A}+\alpha_B v_{0_B}\right)\beta_A \tau_A= d
\eeq

\item for B particles we can also write a condition analogous to equation \eqref{eq:tauA_cond} under the assumption that the motion events for obstacles are determined by encounters among them in absence of mutual interactions; this is justified by the assumption that the concentration of tracers is negligible compared with the  concentration of  obstacles. In this framework is reasonable to assume:
\begin{equation}
\label{eq:tauB_cond}
2\tilde{v}_{0_B}\tilde{\tau}_{B}=2\alpha_B \beta_B \left(v_{0_B}\tau_B\right)=d
\end{equation}

\item the dynamics of tracers is now dominated by the encounters with obstacles and this means that
\begin{equation}
\label{eq:rescCondII}
\dfrac{\tilde{v}_{0_A}^2\tilde{\tau_A}}{3}=D_{exVol_A}(d)
\end{equation}
where $D_{exVol_A}(d)$ is the diffusion coefficient of tracer taking into account the excluded volume effects due to the presence of the obstacles; as we are investigating the case $d \gg R_A+R_B$, we can neglect the excluded volume effects and substitute $D_{0_A}=D_{exVol_A}(\infty)$, yielding:
\beq \label{alfaA2betaA}
\dfrac{\tilde{v}_{0_A}^2\tilde{\tau_A}}{3} = \dfrac{\alpha_A^2 \beta_A \left(v_{0_A}^2\tau_A\right)}{3} = \dfrac{v_{0_A}^2\tau_A}{3}=D_{0_A} \;\;\;\;
\Rightarrow \;\;\;\; \alpha_A^2\beta_A =1 \
\eeq
\item the consideration in the previous item can be extended to obstacles (B-particles) if no interactions act among them, so that:
\begin{equation}
\label{alfaB2betaB}
\dfrac{\tilde{v}_{0_B}^2\tilde{\tau_B}}{3} = \dfrac{\alpha^2_B\beta_B\left(v_{0_B}^2\tau_B\right)}{3} = \dfrac{v_{0_B}^2\tau_B}{3}=D_{0_B} \;\;\;\;
\Rightarrow \;\;\;\; \alpha_B^2\beta_B =1 \ .
\end{equation}
\ei

Notice that the rescaled velocity and time now implicitly depend on the parameter $d$.


Solving the system formed by equations \eqref{eq:tauA_cond}, \eqref{eq:tauB_cond},\eqref{alfaA2betaA} and \eqref{alfaB2betaB}, we obtain:
\begin{equation}
\alpha_B=\dfrac{2\, v_{0_B} \tau_B}{d} \qquad \beta_B=\dfrac{1}{\alpha_B^2}
\end{equation}
while for the rescaled parameters for A-particles: 

\begin{equation}
\alpha_A=\dfrac{v_{0_A}\tau_A}{2d}\left(1\pm\sqrt{1+8\,\dfrac{v_{0_B}^2\tau_B}{v_{0_A}^2\tau_A}}\right) \qquad \beta_A=\dfrac{1}{\alpha_A^2}
\end{equation}

where, as $\alpha_A>0$, the physical solution we choose is the one with "+" sign.

Using equations (\ref{v02_equipart}) and (\ref{eq:taufree}) we can rewrite this as:
\begin{equation}
\alpha_B=\dfrac{2\,\sqrt{3kTm_B}}{\gamma_B d} \qquad \beta_B=\dfrac{1}{\alpha_B^2}
\end{equation}
and
\begin{equation} \label{alfabetaA}
\alpha_A=\dfrac{\sqrt{3kTm_A}}{2\gamma_A d}\left(1 +\sqrt{1+8\,\dfrac{\gamma_A}{\gamma_B}}\right) \qquad \beta_A=\dfrac{1}{\alpha_A^2}
\end{equation}

We suppose that, in the presence of mutually interacting biomolecules of $B$-type, the function $\tilde{\phi}_{0_A}(t)$  is modified as follows
\beq \label{tilphi}
\tilde{\phi}_{A}(t) = q_1\,e^{-t/\tilde{\tau}_A} \;\;\;\;\;\mbox{if}\;\; t<\tilde{T}_d\;,\;\;\;\;\;\;\;\;\;\;
\tilde{\phi}_{A}(t) = q_2\,e^{-t/\tilde{T}_d} \;\;\;\;\;\mbox{if}\;\; t\ge\tilde{T}_d
\eeq
where $q_1,q_2$ are such that $\tilde{\phi}_A(t)$ is normalized and continuous at $t=\tilde{T}_d$. $\tilde{T}_d$ is again the characteristic time at which the motion events described by $\tilde{\psi}_{0_A}({\bf r},t)$ are perturbed by the presence of the obstacles. Equation (\ref{tilphi}) expresses the fact that, on spatial scales larger than the average intermolecular distance $d$ between any pair of obstacles, the timescale of diffusion changes from $\tilde{\tau}_A$ to $\tilde{T}_d$ which is the characteristic time it takes to cover a distance $d$ for a tracer in presence of interacting obstacles.
Two physically equivalent conditions for defining $\tilde{T}_d$ are
\beq \label{tilTd}
\tilde{T}_d \simeq \dfrac{d}{\tilde{v}_{0_A}+\tilde{v}_{0_B}+ V_d}
\eeq

Here $V_d$ is the drift velocity of the obstacles, that we can estimate in the same way as in Section \ref{micro}, that is, $V_d \simeq F(d)/\gamma_B$. For both conditions, it is evident that $\tilde{T}_d\le \tilde{\tau}_A$ where the equality holds when $V_d=0$, that is, the $B$-particles not interact among them.


After a straightforward calculation, we obtain the following dependence of the diffusion coefficient on the parameter
$y(d)= \tilde{T}_d/\tilde{\tau}_A$
\beq \label{DD0_resc}
\dfrac{D_A}{D_{0_A}} = \dfrac{1 - e^{-y} \left( 1+y+\dfrac{y^2}{2}-\dfrac{5y^3}{2} \right)}{1 - e^{-y}\left( 1+y-2y^2\right)}
\eeq

If we take the condition (\ref{tilTd}) for $\tilde{T}_d$ we get for the dependence of $y$ on $d$
\beq \label{Tdtau_resc}
y(d) = \dfrac{\tilde{T}_d}{\tilde{\tau}_A} = \dfrac{1}{1 + \dfrac{V_d\,\tilde{\tau}_A}{d}} = 
\dfrac{1}{1 + \dfrac{F(d)}{d \gamma_B}\beta_A\dfrac{m_A}{\gamma_A}} = \dfrac{1}{1 + \dfrac{4 d \,F(d)\gamma_A}{3 k T\gamma_B  \left(1 +\sqrt{1+8\,\dfrac{\gamma_A}{\gamma_B}}\right)^2}}
\eeq
where we have used equation (\ref{alfabetaA}) for $\beta_A$.

\section{Slowing down of Brownian diffusion: the patterns of $D/D_0$}

In this Section we report the patterns of the ratio $D_A/D_{0_A}$ obtained by means of the theoretical expressions (\ref{DD0}), (\ref{Tdtau}) and (\ref{DD0_resc}), (\ref{Tdtau_resc}). We denote with $D$ and $D_0$  the diffusion coefficients of the tracers ($A$-particles) in the presence and in the absence of obstacles ($B$-particles), respectively. We plot this ratio as a function of the average distance $d$ between any two obstacles obtained for different kinds of interaction potentials between the $B$-particles: screened electrostatic potential, Coulomb potential, dipolar potential. These potentials have been chosen as they are representative of some relevant interaction in biology \cite{stroppolo}. The choice of coulombic and dipolar potentials is justified by the fact that these are long range interactions that can exert their action on a length scale much larger than the typical dimensions of biomolecules. In this framework other interactions, i.e. Van der Waals interactions, have a very short range and they exert their action on length scale comparable with biomolecules dimensions. Nevertheless the short range screened coulombic potential has been investigated as its range distance depends on the free ions concentration in diffusive medium which is an accessible experimental parameter. In what follows the diffusion of tracers in presence of interacting obstacles is studied for some cases corresponding to the different frameworks discussed in Sections \ref{micro}, \ref{rescaled}.

\subsection{Case of modification of the microscopic free-flight time distribution}

As discussed in Section \ref{micro}, this approach corresponds to the case where the characteristic timescale $\tau_A$ of Brownian collisions is of the same order of magnitude than $T_d$ (the characteristic timescale at which the tracers $A$ "see" the obstacles $B$). This corresponds to intermolecular distances $d$ of the obstacles that are comparable to $\sqrt{\dfrac{m_A k T}{\gamma_A^2}}\,$. For  the  sake of simplicity we consider the case for which the species $A$ and $B$ have the same size, $R=R_A=R_B$, and the same mass, $m=m_A=m_B$, which define a length and a mass scale for the system, respectively. Hence, for instance, the distance between two colliding particles can be rewritten as $d=R\,l$,  where $l$ is an adimensional parameter,  with the assumption that $d\gg R$. Moreover, the temperature $T$ of the system defines an energy scale allowing to express equation (\ref{Tdtau}) in terms of adimensional parameters,  since  the friction coefficient as well can be expressed in terms of an  adimensional parameter $\Gamma$
\begin{equation}
\gamma=\Gamma(k T m)^{1/2}R^{-1} 
\end{equation}
Let us consider a two-body interaction potential of the form
\begin{equation}
\label{eq:invmonPot}
U(r)=\dfrac{\mathcal{C}}{r^{n}}
\end{equation}
where $r$ is the interparticle distance, which can be written in adimensional units as

\begin{equation}
\label{eq:invmonPot_norm}
U(r=Rl)=\bar{U}(l)=(k T) \bar{\mathcal{C}} l^{-n}
\end{equation}
where $\bar{\mathcal{C}}=\mathcal{C} (k T R^{n})^{-1}$.
With these conventions Equation \eqref{Tdtau} reads
\begin{equation}
\label{eq:xforgraph}
x=x(d=Rl)=\dfrac{T_d}{\tau}=\dfrac{l\Gamma^2}{2\sqrt{3}\Gamma +\bar{\mathcal{C}} n l^{-(n+1)}}
\end{equation}

Let us consider the case of a coulombic interaction
\begin{equation}
U_{Coul}(r)=\mathcal{C}_{Coul}r^{-1}
\end{equation}
among the $B$-type particles. In order to study a somewhat realistic case we  take  for $m$ and $R$ values that are typical for macromolecules , i.e. $m\sim 10\,\text{KDa}\simeq 1.6\times 10^{-23}$Kg and $R\simeq 10^{-9}$m and $|Z|\simeq 10$, at room temperature $T=300 \, K$; in such a case we have
\begin{equation}
\label{eq:aCoul}
\bar{\mathcal{C}}_{Coul}=\dfrac{Z^2 q^2}{\varepsilon_{water} (kTR)}\simeq 0.7\times 10^{2}
\end{equation}
where $q$ is the electric elementary charge and $\varepsilon_{water}\simeq 80$  is the relative electric permittivity of water.

\begin{figure}[h!]
 \centering
 \includegraphics[scale=0.5,keepaspectratio=true]{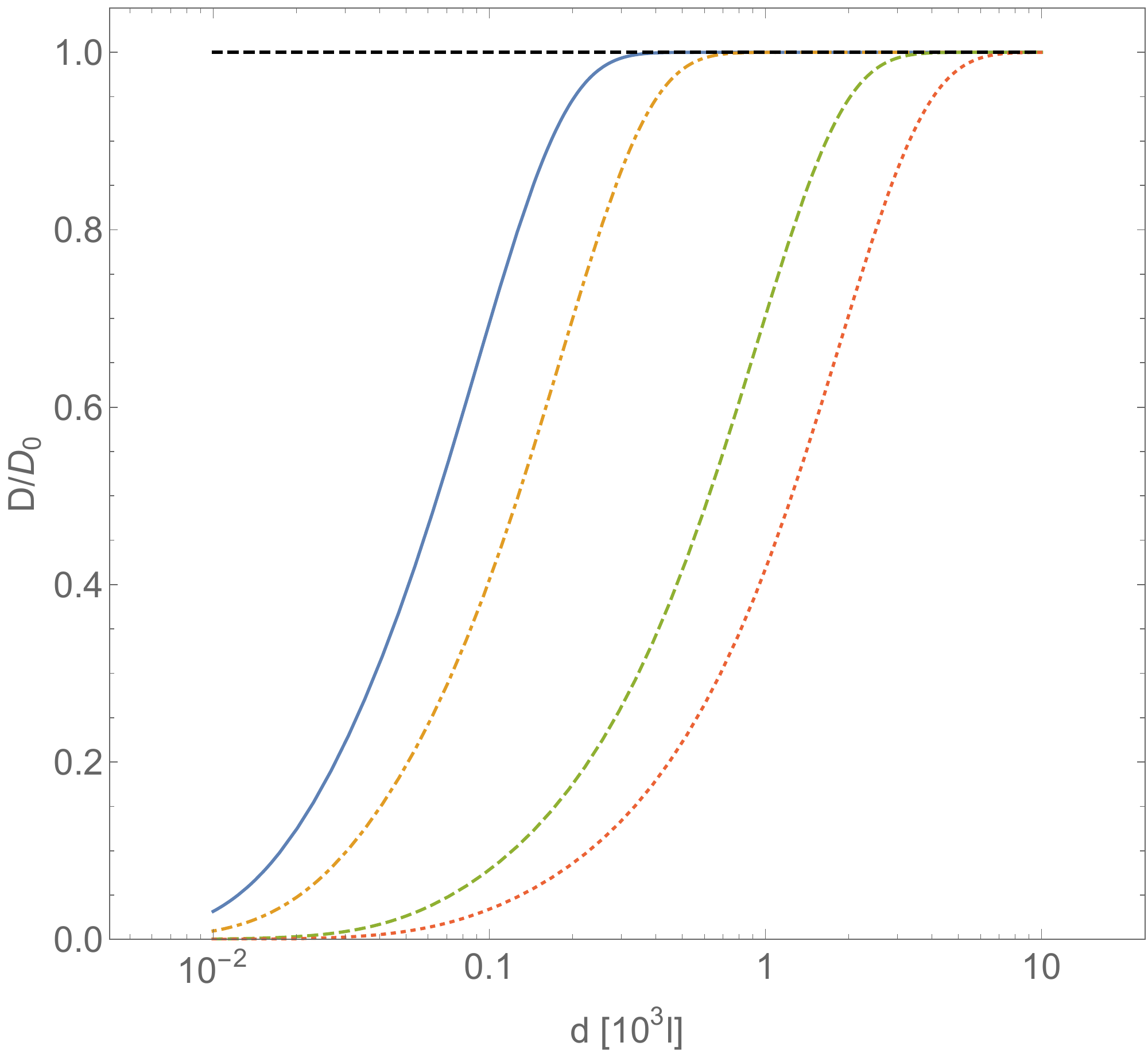}
 \caption{ \label{fig:FiDrelCoulfixedalpha} Normalized diffusion coefficient $D/D_0$ for $A$-type particles, computed with Eqs. \eqref{DD0} and \eqref{Tdtau}, plotted vs. the intermolecular average distance $d$ of $B$-type particles (expressed in adimensional units $l$). The $B$-particles interact through a coulombic potential $U=\bar{\mathcal{C}}_{Coul}l^{-1}$. The $A$- and $B$-type particles are assumed spherical, of equal radius $R$, and equal mass $m$.  
In adimensional units the interaction intensity is $\bar{\mathcal{C}}_{Coul}=U_{Coul}(R)/(k_{B}T)$, the friction coefficient  $\gamma=\Gamma (k_{B}Tm)^{1/2} R^{-1}$. The curves refer to a fixed value for the potential strength ($\bar{\mathcal{C}}_{Coul}=70$) and different values for $\Gamma$, that is: $\Gamma=0.1$ (continuous line), $\Gamma=0.05$ (dot-dashed line), $\Gamma=0.01$ (dashed line), $\Gamma=0.005$ (dotted line).}
\end{figure}

In Figure \ref{fig:FiDrelCoulfixedalpha}  we plot the tracer self-diffusion coefficent behavior as a function of average distances among diffusing obstacles interacting through a coulombic potential, following equations \eqref{DD0} and \eqref{eq:xforgraph}; the intensity of coulombic potential has been fixed to $\bar{\mathcal{C}}_{Coul}=0.7\times 10^{2}$ while the value of the adimensionalized friction coefficient $\Gamma$ has been changed. In this case it is necessary to choose  $\Gamma\simeq 10^{-2}$ in order to obtain sizable effects on the value of $D/D_{0}$ at an average intermolecular distance of about $l\simeq 10^{3}$. Moreover the value of $\Gamma$ strongly affects the value of the intermolecular average distance among obstacles which corresponds to a major deviation of the tracer self diffusion coefficient from its Brownian value: the smaller the value of $\Gamma$ is, the larger the distance among obstacles at which diffusion of tracers deviates from Brownian diffusion.
Assuming that the friction coefficient is given by the Stokes' law \eqref{eq:StokesLaw}, the $\Gamma$ value obtained corresponds to $\eta\simeq 1.5\times 10^{-4}\eta_{water}$, where $\eta_{water}$ is the viscosity of water at the temperature $T=300 \,K$.

\begin{figure}[h!]
 \centering
 \includegraphics[scale=0.5,keepaspectratio=true]{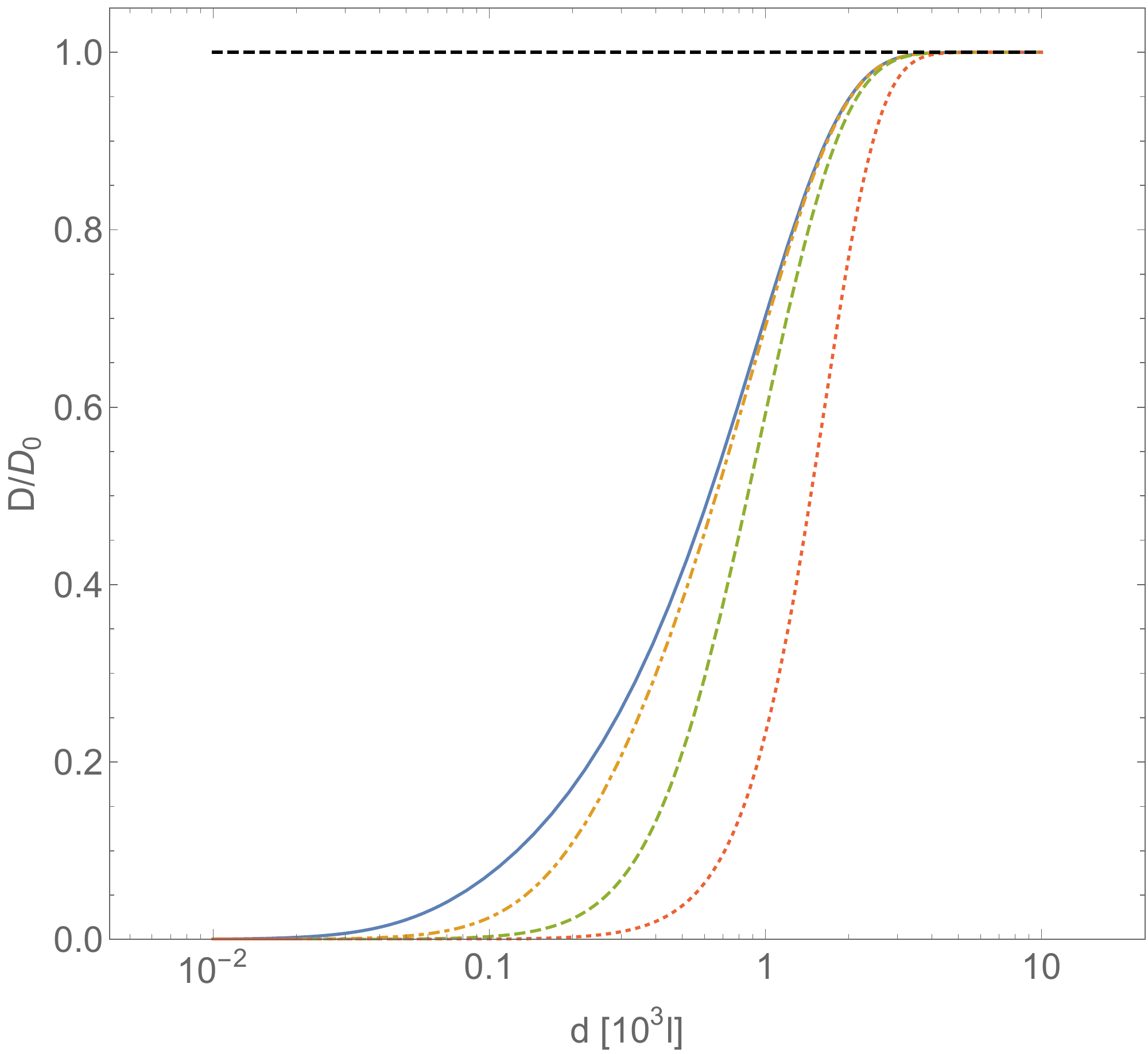}
 \caption{\label{fig:FigDrelCoulfixedgamma} Normalized diffusion coefficient $D/D_0$, computed with Eqs. \eqref{DD0} and \eqref{Tdtau}, for $A$-type particles vs. intermolecular average distance $d$ of $B$-type particles (expressed in adimensional units). The $B$-particles interact through a coulombic potential $U_{Coul}=\bar{\mathcal{C}}l^{-1}$. Conventions on units are the same of Fig.\ref{fig:FiDrelCoulfixedalpha}. The curves refer to the fixed value $\Gamma=0.01$ of the friction coefficient, and to different values of the potential strength: $\,\bar{\mathcal{C}}_{Coul}=10^{2}$ (continuous line), $\,\bar{\mathcal{C}}_{Coul}=10^{3}$ (dot-dashed line), $\,\bar{\mathcal{C}}_{Coul}=10^{4}$ (dashed line), $\,\bar{\mathcal{C}}_{Coul}=10^{5}$ (dotted line).}
\end{figure}

In Figure \ref{fig:FigDrelCoulfixedgamma} we plot the tracer self-diffusion coefficent behavior as a function of average distances among diffusing obstacles interacting through a coulombic potential, for a fixed value of  $\,\Gamma=10^{-2}$ and different values for the strength of coulombic interaction among obstacles. In this case we observe that as we increase the strength of coulombic potential among obstacles the profile of tracer self-diffusion coefficient as a function of the average distance among obstacles becomes sharper. Nevertheless, the intensity of the potential does not seem to affect the value of average distance among obstacles at which the tracers self diffusion coefficient deviates from its gaussian value.

As mentioned above, the renormalized self-diffusion coefficient of tracers has been computed in presence of obstacles interacting through a ``dipole-dipole'' potential 
\begin{equation}
U_{Dip}(r)=\mathcal{C}_{Dip} r^{-3} 
\end{equation}
and a screened coulombic potential, of a form close to the Debye-H\"{u}ckel potential (which usually models electrostatic interactions in electrolytic solutions), that is
\begin{equation}
\label{eq:DebyePot}
 U_{CoulScr}(r)=\dfrac{\mathcal{C}_{CoulScr}\exp\left[ -r/\lambda_D \right]}{r}
\end{equation}
where $\lambda_D$ is the characteristic screening length scale, also called Debye length.

In adimensional form the potential in \eqref{eq:DebyePot} is rewritten as

\begin{equation}
U_{CoulScr}(r=Rl)=\bar{U}_{CoulScr}(l)=\dfrac{U_{Coul}(R)}{k T}\dfrac{\exp\left[-\dfrac{l-1}{\bar{\lambda}_D}\right]}{l}=\bar{\mathcal{C}}_{CoulScr}\dfrac{\exp\left[-\dfrac{l-1}{\bar{\lambda}_D}\right]}{l}
\end{equation}
where $\bar{\lambda}_D=\lambda_D/R$ is the adimensional screening length. As pointed out above, the method proposed in the present paper is meaningful provided that $d \gg R$, therefore we take $\,\bar{\lambda}_D \geq 10 $ because for shorter screening length scales we don't expect any effect of the interactions among the obstacles on the diffusion of the tracers. For the screened coulomb potential equation \eqref{Tdtau} takes the form
\begin{equation}
\label{eq:xforgraph_CoulScreen}
x=x(d=Rl)=\dfrac{T_d}{\tau_A}=\dfrac{l\Gamma^2}{2\sqrt{3}\Gamma +\bar{\mathcal{C}}_{CoulScr} \left(\dfrac{1}{l^{2}}+\dfrac{1}{l\bar{\lambda}_{D}}\right)\exp\left(-\dfrac{l-1}{\bar{\lambda}_D}\right)}
\end{equation}

\begin{figure}[h!]
 \centering
 \includegraphics[scale=0.5,keepaspectratio=true]{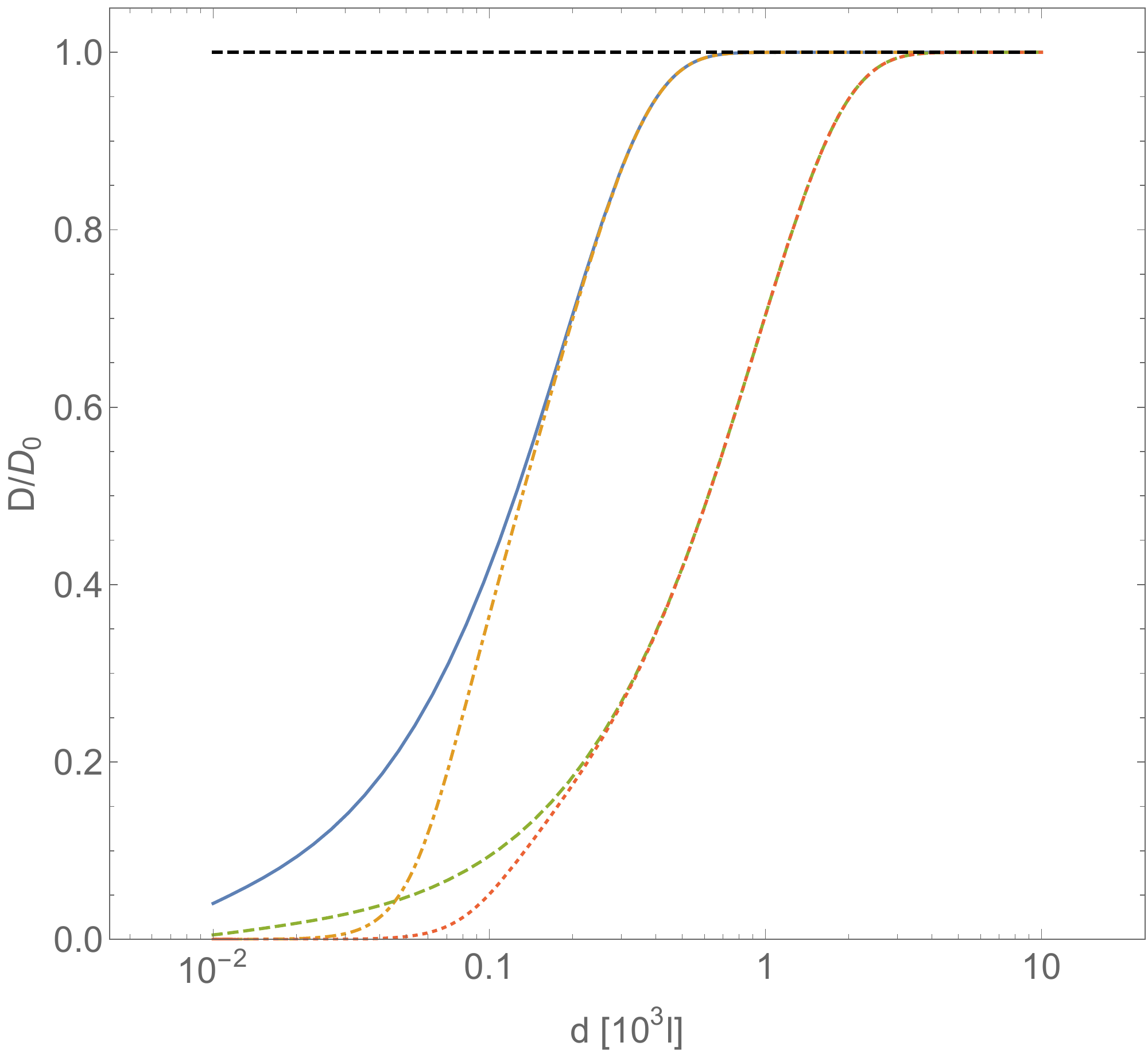}
 \caption{\label{fig:FigDrelCub} Normalized diffusion coefficient $D/D_0$, computed with Eqs. \eqref{DD0} and \eqref{Tdtau}, for $A$-type particles vs.intermolecular average distance $d$ of $B$-type particles (expressed in adimensional units). The $B$-type particles interact through a dipolar potential $U(r)=\mathcal{C}_{Dip} r^{-3}=\bar{\mathcal{C}}_{Dip}(r/R)^{-3}$. Conventions on adimensional units are the same of Fig.\ref{fig:FiDrelCoulfixedalpha}. The curves refer to different choices of the friction coefficient $\Gamma$ and of the  strength $\bar{\mathcal{C}}_{Dip}$ of the potential energy: $\Gamma=0.05 \,\,\text{and}\,\, \bar{\mathcal{C}}=10^{2}$ (continuous line), $\Gamma=0.05\,\,\text{and}\,\,\bar{\mathcal{C}}=10^{6}$ (dot-dashed line), $\Gamma=0.01\,\,\text{and}\,\,\bar{\mathcal{C}}=10^{2}$ (dashed line), $\Gamma=0.01\,\,\text{and}\,\,\bar{\mathcal{C}}=10^{6}$ (dotted line).}
\end{figure}

\begin{figure}[h!]
 \centering
 \includegraphics[scale=0.5,keepaspectratio=true]{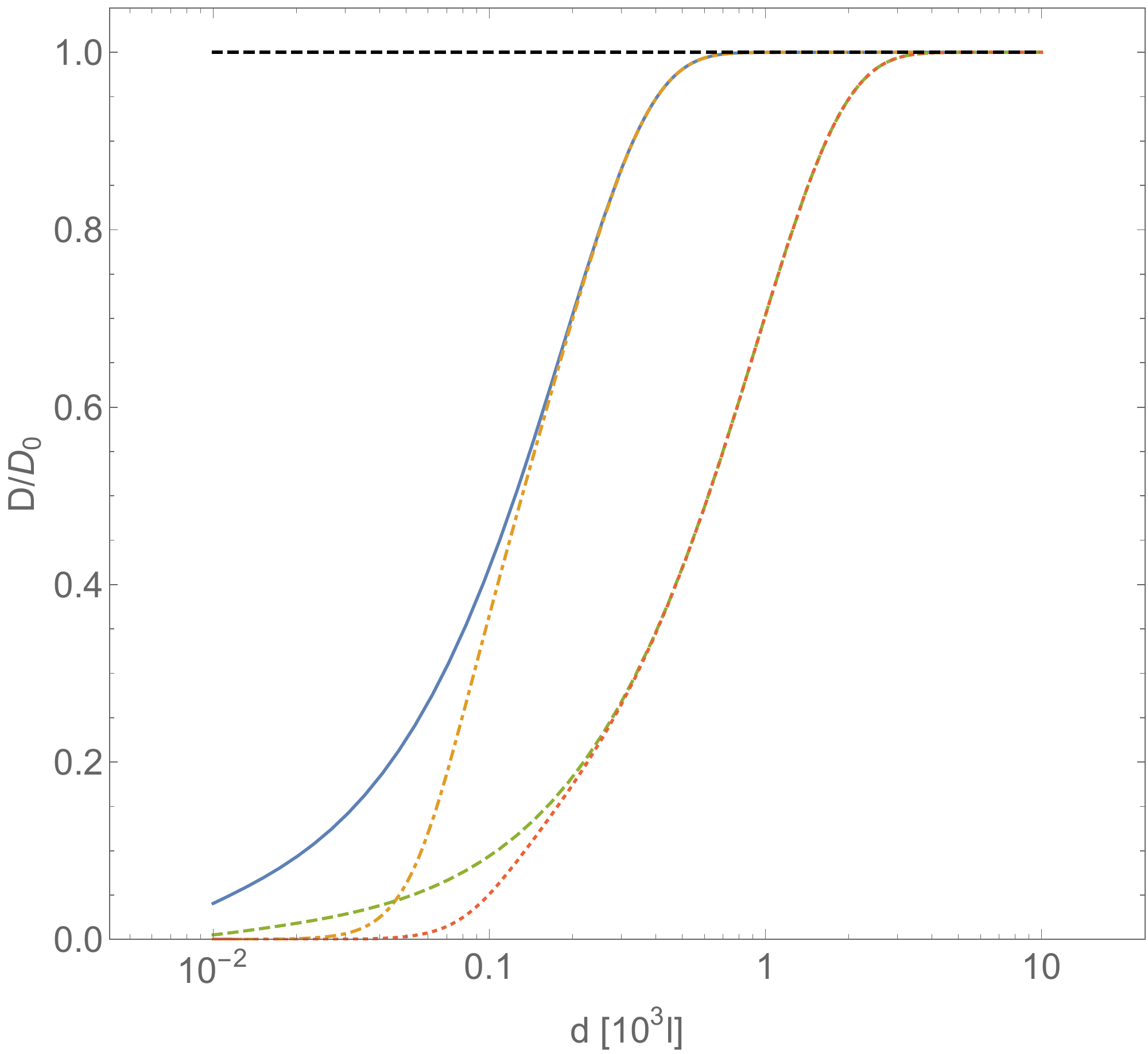}
 \caption{\label{fig:FigDrelDebye} Normalized diffusion coefficient $D/D_0$, computed with Eqs. \eqref{DD0} and \eqref{Tdtau}, for $A$-type particles vs. intermolecular average distance $d$ of $B$-type particles (expressed in adimensional units). The $B$-type particles interact through the coulombic screened potential given in \eqref{eq:DebyePot}. Conventions on adimensional units are the same of Fig.\ref{fig:FiDrelCoulfixedalpha}. The curves refer to different choices of the value of the friction coefficient  $\Gamma$ and of the screening length $\lambda_D=10$ which set the strength of the potential energy: $\,\Gamma=0.05 \,\,\bar{\mathcal{C}}_{CoulScr}=10^{2}$ (continuous line), $\,\Gamma=0.05\,\,\bar{\mathcal{C}}_{CoulScr}=10^{6}$ (dot-dashed line), $\,\Gamma=0.01\,\,\bar{\mathcal{C}}=10^{2}$ (dashed line), $\,\Gamma=0.01\,\,\bar{\mathcal{C}}_{CoulScr}=10^{6}$ (dotted line).}
\end{figure}

In Figures \ref{fig:FigDrelCub} and \ref{fig:FigDrelDebye} we show the behavior of tracers self-diffusion coefficient as a function of concentration of interacting obstacles, in the case of "dipolar" interaction and Coulomb screened interaction among obstacles, respectively. Different values for $\Gamma$, $\bar{\mathcal{C}}_{Dip}$ and $\bar{\mathcal{C}}_{CoulScr}$ have been chosen. In both cases we observe that the dependence of the tracers self-diffusion coefficient on the concentration of obstacles is much more affected by the value of $\Gamma$ then by the strength of the interaction potential among obstacles $\bar{\mathcal{C}}_{Dip}$ and $\bar{\mathcal{C}}_{CoulScr}$ in the explored range of parameters.



\subsection{Case of modification of the rescaled free-flight time distribution}

\begin{figure}[h!]
 \centering
 \includegraphics[scale=0.5,keepaspectratio=true]{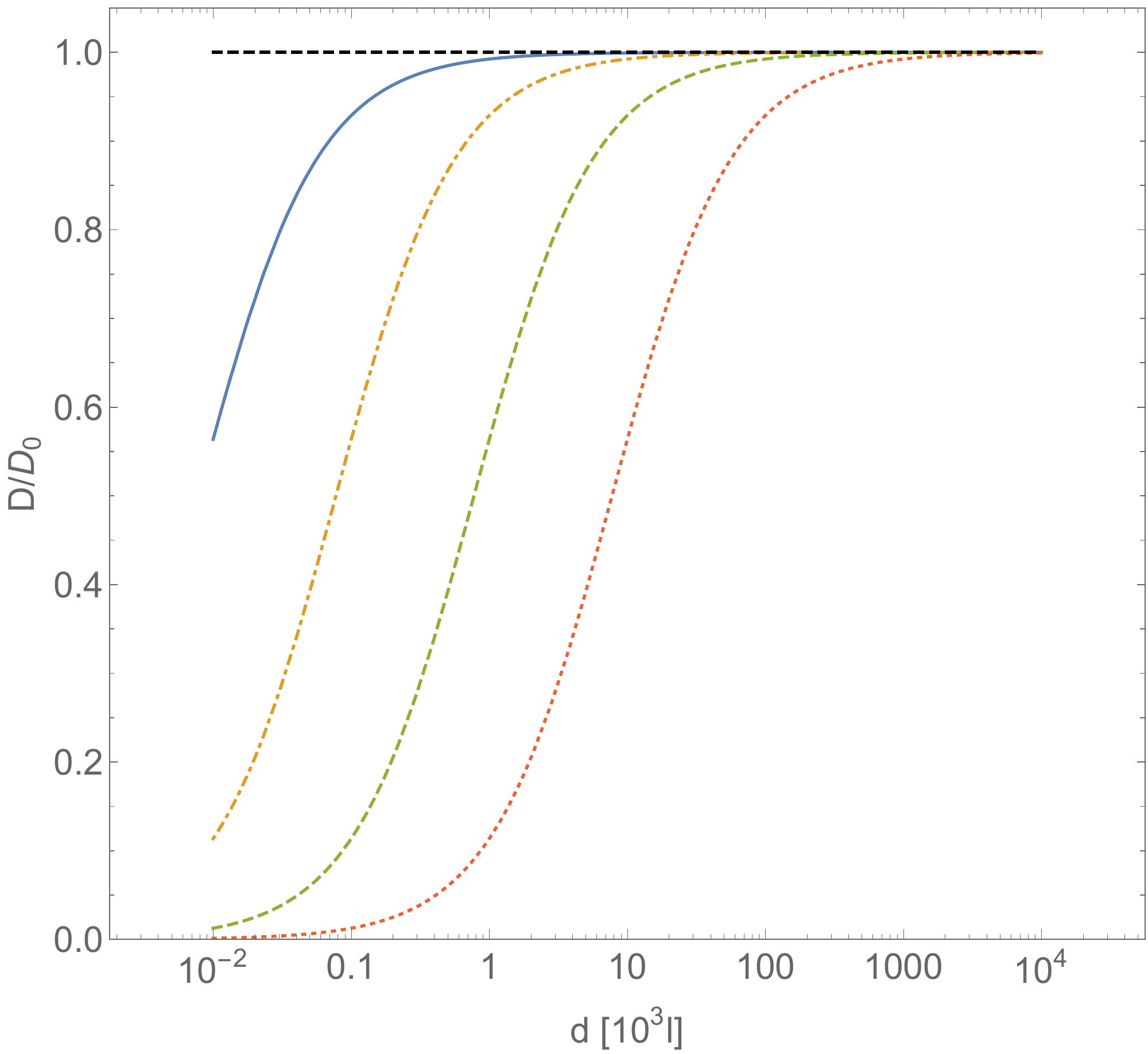}
 \caption{\label{fig:FigDrelCoulresc} Normalized diffusion coefficient $D/D_0$, computed with Eqs. \eqref{DD0_resc} and \eqref{Tdtau_resc}, for $A$-type particles vs. intermolecular average distance $d$ of $B$-type particles (expressed in adimensional units). The $B$-type particles interact through the coulombic  potential $U=\bar{\mathcal{C}}l^{-1}$. Conventions on adimentional units are the same of Fig.\ref{fig:FiDrelCoulfixedalpha}. The curves refer to different values of potential strength: $\bar{\mathcal{C}}_{Coul}=10^{2}$ (continuous line), $\,\bar{\mathcal{C}}_{Coul}=10^{3}$ (dot-dashed line), $\,\bar{\mathcal{C}}_{Coul}=10^{4}$ (dashed line), $\,\bar{\mathcal{C}}_{Coul}=10^{5}$ (dotted line).}
\end{figure}

\begin{figure}[h!]
 \centering
 \includegraphics[scale=0.5,keepaspectratio=true]{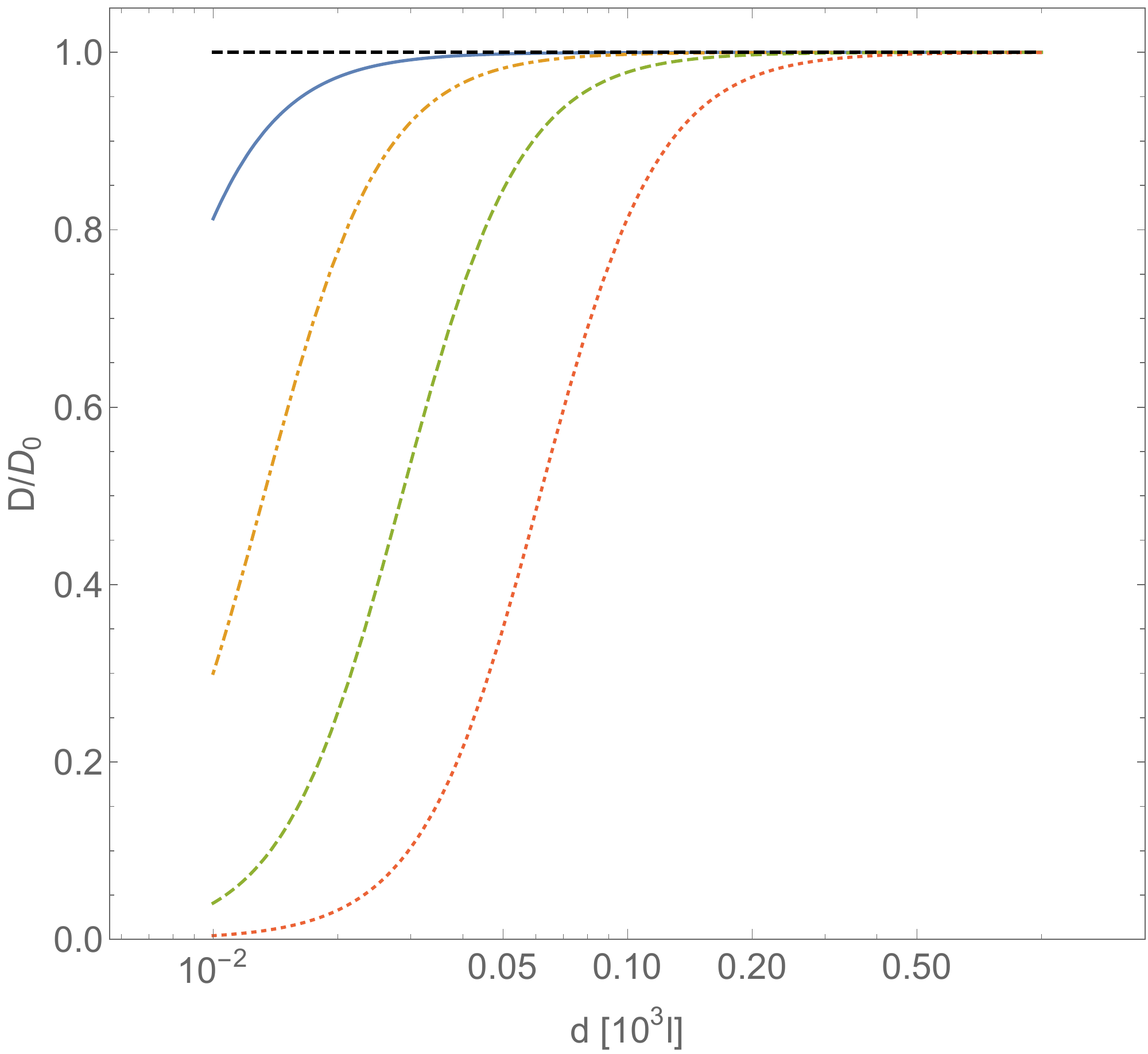}
 \caption{\label{fig:FigDrelCubresc}Normalized diffusion coefficient $D/D_0$, computed with Eqs. \eqref{DD0_resc} and \eqref{Tdtau_resc}, for $A$-type particles vs. intermolecular average distance $d$ of $B$-type particles (expressed in adimensional units). The $B$-type particles interact through a dipolar  potential $U=\bar{\mathcal{C}}l^{-3}$. Conventions on adimensional units are the same of Fig \ref{fig:FiDrelCoulfixedalpha}. The curves refer to  different values for potential strength: $\,\bar{\mathcal{C}}_{Dip}=10^{2}$ (continuous line), $\,\bar{\mathcal{C}}_{Dip}=10^{3}$ (dot-dashed line), $\,\bar{\mathcal{C}}_{Dip}=10^{4}$ (dashed line), $\,\bar{\mathcal{C}}_{Dip}=10^{5}$ (dotted line).}
\end{figure}

As discussed in Section \ref{rescaled}, the proposed approach corresponds to the case where the characteristic timescale $\tau$ of Brownian collisions much smaller than the transition time $T_d$. This corresponds to intermolecular distances $d$ of the obstacles that are much larger than $\sqrt{\dfrac{m_A k T}{\gamma_A^2}}\,$. We remark that if $\gamma$ is given by the Stokes law \eqref{eq:StokesLaw} then the  collision time $T_d$ does not depend on the viscosity of the medium surrounding the particles but only on the ratio between the radii of the $A$- and $B$-type particles, on the functional form of the interaction potential between the obstacles, and on the strength of this potential. As in the previous section, we choose identical $A$- and $B$-particles in order to introduce adimensional units. For a potential of the form $U=\mathcal{C} r^{-n}$ Equation \eqref{Tdtau_resc} is rewritten as follows

\begin{equation}
y(d=Rl) = \dfrac{1}{1 + \dfrac{4 d \,F(d)\gamma_A}{3 k T\gamma_B  \left(1 +\sqrt{1+8\,\dfrac{\gamma_A}{\gamma_B}}\right)^2}} = \dfrac{1}{1+\dfrac{1}{12}n\bar{\mathcal{C}}l^{-n}}
\end{equation}

as $\gamma_A=\gamma_B$. In Figures \ref{fig:FigDrelCoulresc} and \ref{fig:FigDrelCubresc} we report the different patterns obtained for $D/D_0$ of the tracers ($A$-particles) as a function of the average distance $d$ between any pair of obstacles ($B$-particles) interacting through the coulombic and dipolar potential. 

\section{Discussion}
Let us now make some remarks about the results reported in the present paper.
First of all notice that, under the assumptions made to model through the CTRW approach the diffusion behavior of a random walker in an environment crowded by randomly moving obstacles, the random walker (also called "tracer" throughout the paper) still makes a Brownian diffusion. No anomalous diffusion law is found. Instead of the diffusion law it is rather the value of the Brownian diffusion coefficient which is affected by the randomly moving obstacles. The fact that the obstacles move under the influence of deterministic nonlinear interparticle potentials  implies a chaotic dynamics which a-priori could be very different from a stochastic dynamics, this notwithstanding such a chaotic dynamics entails a Brownian-like diffusion as was found by numerical simulations in Ref.\cite{pre2}. This circumstance makes it reasonable to represent the dynamics of the interacting obstacles through an effective dynamics resulting from a sequence of random displacements as is assumed throughout the present work.

We have found that the reduction of the value of the diffusion coefficient of the tracers can be very large in presence of interacting obstacles, and this fact can have relevant consequences for several applications. In particular, the description of the complex network of biochemical reactions taking place in living cells could be markedly affected by the activation of long-range intermolecular interactions of the kind discussed in Ref.\cite{pre3}. For instance, if we imagine a cytoplasm crowded by biomolecules interacting at a long distance then those molecules that would be driven to their targets only by diffusion could be considerably slowed  down. Many other dynamical scenarios are possible stemming from the interplay of standard and chaotic diffusion, and the dynamical crowding investigated above.

\begin{acknowledgments}
The authors wish to thank F. Piazza and R. Lima for useful comments and suggestions.
This work	was supported by the Seventh Framework Programme for Research of the European Commission under FET-Open grant TOPDRIM (Grant No. FP7-ICT-318121).
\end{acknowledgments}

\section{Appendix}

In this Section, we compute the probability distribution $P({\bf r}, t)$ for the walker to be at location ${\bf r}$, at time $t$, following \cite{ZK93} and generalising the result to the three-dimensional case.

Let $\psi({\bf r}, t)$ be, as in section \ref{ctrw}, the probability density of making a displacement ${\bf r}$ in time $t$ in a single motion event:
$$
\psi({\bf r}, t) = \Lambda({\bf r})\; \delta\left(t - \dfrac{\left|{\bf r}\right|}{v_0}\right)
$$
The probability $Q({\bf r}, t)$ of arriving at location ${\bf r}$ exactly at time $t$ and to stop before randomly choosing a new direction satisfies the recursion relation:
$$
Q({\bf r}, t) = \int_0^{+\infty} dt' \int d^3{\bf r'} \;Q({\bf r}-{\bf r'}, t-t')\;\psi({\bf r'}, t') + \delta({\bf r})\delta(t)
$$

\subsection{Jump Model}

In the Jump Model, particles wait at a particular location before moving instantaneously to the next one, the displacement being chosen according to the probability density $\Lambda({\bf r})$, the waiting time before the jump is $|{\bf r}|/v_0$ [because of the $\delta$-function in the expression of $\psi({\bf r}, t)$].

The three-dimensional formulation is straigthforward in this case (and it appears for example in \cite{KBS87}).  
We have for the probability distribution $P({\bf r}, t)$:
$$
P({\bf r}, t) = \int_0^{t} dt' \;Q({\bf r}, t-t')\;\Psi(t')
$$
where $\Psi(t)$ is the probability for not leaving a position up to time $t$:
$$
\Psi(t) = \int_t^{+\infty} dt' \int d^3{\bf r} \;\psi({\bf r}, t')
$$
Passing to Fourier-Laplace transform defined by:
$$
f({\bf k}, s) = \int_0^{+\infty} dt\,e^{-st} \int d^3{\bf r}\,e^{i {\bf k}\cdot{\bf r}} \,f({\bf r},t)
$$
we get
$$
Q({\bf k}, s) = \dfrac{1}{1 - \psi({\bf k},s)}
$$
so that
$$
P({\bf k}, s) = \dfrac{\Psi(s)}{1 - \psi({\bf k},s)}
$$
The mean square displacement $\langle r^2(t)\rangle$ is the inverse Laplace transform of the quantity
\beq \label{r2s}
\langle r^2(s)\rangle = -\,\left.\Delta_{{\bf k}} P({\bf k}, s)\right|_{{\bf k}=0} =
-\,\left.\dfrac{\Psi(s)}{(1 - \psi({\bf k},s))^2}\left[ \Delta_{{\bf k}}\psi({\bf k},s) +  
\dfrac{2}{1 - \psi({\bf k},s)} (\nabla_{{\bf k}}\psi({\bf k},s))^2 \right]\right|_{{\bf k}=0}
\eeq
where $\Delta_{{\bf k}}$ is  the Laplacian ($\Delta_{{\bf k}} =
\partial^2/\partial k_x^2 + \partial^2/\partial k_y^2 + \partial^2/\partial k_z^2$)
 and $\nabla_{{\bf k}}$ is the gradient ($\nabla_{{\bf k}} = (\partial/\partial k_x\,,\partial/\partial k_y\,,\partial/\partial k_z)$).

We now use the fact that in our case diffusion is isotropic. As discussed in Section \ref{ctrw}, this allows to write
$$
\psi({\bf r},t) = \dfrac{\phi(t)}{4\pi\,(v_0 t)^2}\,\delta(\left|{\bf r}\right| - v_0 t)
$$
where we have introduced the waiting time distribution $\phi(t)$, which is the probability density function that a single motion event has duration $t$, and is normalised by $\int_0^{+\infty} dt\,\phi(t) = 1$. It is easy to show that
\beq \label{Psi_s}
\Psi(s) = \dfrac{1 - \phi(s)}{s}\;,
\eeq
\beq \label{psi_k0s}
\psi({\bf k}=0,s) = \phi(s)\;,
\eeq
\begin{equation}
\label{lapl_psi_k0s}
 \Delta_{{\bf k}}\psi({\bf k},s)|_{{\bf k}=0}=-v_0^2\dfrac{\mathrm{d}^2}{\mathrm{d}s^2}\phi(s)\;,
\end{equation}
where $\phi(s)$ is the Laplace transform of $\phi(t)$.

Isotropy implies that $\left.\nabla_{{\bf k}}\psi({\bf k},s)\right|_{{\bf k}=0} = 0$, so that, replacing equations (\ref{Psi_s}), (\ref{psi_k0s}), (\ref{lapl_psi_k0s}) in equation (\ref{r2s}) we get
\beq \label{r2s_simpl}
\langle r^2(s)\rangle = \dfrac{v_0^2}{(1 - \phi(s))\,s}\cdot\dfrac{\mathrm{d}^2}{\mathrm{d}s^2}\phi(s)
\eeq

Expanding expression (\ref{r2s_simpl}) around $s\simeq 0$, we obtain
$$
\langle r^2(s)\rangle \simeq\dfrac{v_0^2 \langle t^2\rangle_{\phi}}{s^2\langle t \rangle_{\phi}}
$$
Using Tauberian theorems \cite{Fel71}, which relate the behavior of a function $f(t)$ at large $t$ to that of its Laplace transform at small $s$, we have at large times:
$$
\langle r^2(t)\rangle \simeq \dfrac{v_0^2\langle t^2\rangle_{\phi}}{\langle t\rangle_{\phi}}\;t
$$
which is the same as equation (\ref{r2t}) of section \ref{ctrw}.

\subsection{Velocity Model}

In the Velocity Model, each walker moves with constant velocity $v_0$ between turning points where a new direction and a new distance of flight are chosen according to the probability density $\Lambda({\bf r})$. We have in this case:
$$
P({\bf r}, t) = \int_0^{t} \mathrm{d}t' \int\mathrm{d}^3{\bf r'}\;Q({\bf r}-{\bf r}', t-t')\;\Psi(\mathbf{r}',t')
$$
where $\Psi(\mathbf{r},t)$ represents the probability for a particle to make a displacement $\mathbf{r}$ in a time $t$ in a single motion event and without stopping at time $t$. The explicit expression for $\Psi(\mathbf{r},t)$ in 3 dimensions is given by:
$$
 \Psi({\bf r},t)=p_{\alpha,\beta}(r|t)\int \mathrm{d}^3 {\bf r}'\int_0^{+\infty}\mathrm{d}t'\;\psi({\bf r}',t')\theta(\left|{\bf r}'\right|-\left|{\bf r}\right|)\theta(t'-t)\delta(\alpha'-\alpha)\delta(\beta'-\beta)
$$
where $\alpha,\alpha',\beta,\beta'$ are the angles which define the direction of vectors ${\bf r}$ and ${\bf r}'$ in a polar reference system and $p_{\alpha,\beta}(r|t)$ is the conditional probability of making a displacement of distance $r$ in a time interval $t$ along a vector whose orientation is specified by angles $\alpha$ and $\beta$. Heaviside functions $\theta(x)$ take into account time ordering $t'>t$ so that $|{\bf r}'|-|{\bf r}|>0$, as the velocity is constant.

We again consider the Fourier-Laplace transform of the previous functions, obtaining:
$$
Q({\bf k}, s) = \dfrac{1}{1 - \psi({\bf k},s)}
$$
and
$$
P({\bf k}, s) = \dfrac{\Psi({\bf k},s)}{1 - \psi({\bf k},s)}.
$$
The mean square displacement $\langle r^2(t)\rangle$ as a fuction of time is the inverse Laplace transform of the quantity:
\begin{equation}
\begin{split}
\label{eq:meansquaredisp}
& \langle r^2(s)\rangle = -\Delta_{{\bf k}=0}P({\bf k},s)|_{{\bf k}=0}=-\biggr[\dfrac{\Delta_{{\bf k}} \Psi({\bf k},s)}{(1-\psi({\bf k},s))}+
\dfrac{2\nabla_{{\bf k}}\psi({\bf k},s)\cdot\nabla_{{\bf k}}\Psi({\bf k},s)}{(1-\psi({\bf k},s))^2}\\
\\
& +\dfrac{2\Psi({\bf k},s)|\nabla_{{\bf k}}\psi({\bf k},s)|^2}{(1-\psi({\bf k},s))^3}+\dfrac{\Psi({\bf k},s)\Delta_{{\bf k}}\psi({\bf k},s)}{(1-\psi({\bf k},s))^2}
\biggr]\biggr|_{{\bf k}=0}
\end{split}
\end{equation}

As we consider the isotropic case, we can rewrite $\psi({\bf r},t)$ as
$$
\psi({\bf r},t) = \dfrac{\phi(t)}{4\pi\,(v_0 t)^2}\,\delta(\left|{\bf r}\right| - v_0 t) \ ,
$$
under such hypothesis $\Psi({\bf r},t)$ has the form:
\beq \nonumber
\begin{split}
 \Psi({\bf r},t)=&\dfrac{\delta(|{\bf r}|-v_0 t)}{v_0^2 t^2}\int \mathrm{d}^3 {\bf r}'\int_0^{+\infty}\mathrm{d}t' \;\dfrac{\phi(t')}{4 \pi v_0^2 t'^2}\delta(|{\bf r}'| - v_0 t')\\ 
 \\
 &\times \theta(\left|{\bf r}'\right|-\left|{\bf r}\right|)\theta(t'-t)\delta(\alpha'-\alpha)\delta(\beta'-\beta)\ . \\
 \end{split}
\eeq

The isotropy hypothesis implies $\left.\nabla_{{\bf k}}\psi({\bf k},s)\right|_{{\bf k}=0} = 0$. Equation (\ref{eq:meansquaredisp}) then reduces to:
\begin{equation}
\begin{split}
\label{eq:meansquaredispsimply}
& \langle r^2(s)\rangle = -\dfrac{1}{(1-\psi({\bf k},s))}\biggr[\Delta_{{\bf k}}\Psi({\bf k},s)
+\dfrac{\Psi({\bf k},s)\Delta_{{\bf k}}\psi({\bf k},s)}{(1-\psi({\bf k},s))}
\biggr]\biggr|_{{\bf k}=0} \ .
\end{split}
\end{equation}

It is easy to show that:
\begin{equation}
\label{eq:Psitrasf}
\Psi({\bf k}=0,s)=\dfrac{1-\phi(s)}{s}
\end{equation}
and
\begin{equation}
\label{eq:laplPsitrasf}
 \Delta_{{\bf k}}\Psi({\bf k},s)|_{{\bf k}=0}=-\,v_0^2\,\dfrac{\mathrm{d}^2}{\mathrm{d}s^2}\biggr[\dfrac{1-\phi(s)}{s}\biggr]
\end{equation}
where $\phi(s)$ is the Laplace transform of $\phi(t)$.

Replacing the Fourier-Laplace transforms (\ref{psi_k0s}),(\ref{lapl_psi_k0s}),(\ref{eq:Psitrasf}), (\ref{eq:laplPsitrasf}) in equation (\ref{eq:meansquaredispsimply})
we obtain:
\begin{equation}
 \begin{split}
\label{eq:meansquarelapltrans}
& \langle r^2(s)\rangle = \dfrac{v_0^2}{(1-\phi(s))}\biggr[\dfrac{\mathrm{d}^2}{\mathrm{d}s^2}\biggr(\dfrac{1-\phi(s)}{s}\biggr)
+\dfrac{1}{s}\dfrac{\mathrm{d}^2}{\mathrm{d}s^2}\phi(s)\biggr]
\end{split}
\end{equation}


Expanding expression (\ref{eq:meansquarelapltrans}) around zero, we obtain:
$$
\langle r^2(s)\rangle \simeq\dfrac{v_0^2 \langle t^2\rangle_{\phi}}{s^2\langle t \rangle_{\phi}}
$$
and using Tauberian theorems \cite{Fel71}, we have at large times:
$$
\langle r^2(t)\rangle \simeq\dfrac{v_0^2 \langle t^2\rangle_{\phi}}{\langle t \rangle_{\phi}}t
$$
as in the Jump Model.

\section{Competing interests}
The authors declare that they have no competing interests.

\section{Authors' contributions}
MG and EF developed the application of the Continuous Time Time Random Walk technique to the special problem considered. IN and ID  contributed to the numerical analysis and to the discussions defining the model. MP proposed the problem and supervised the overall development of the work. All the authors participated in the scientific discussions and to the writing and editing of the paper.

\end{document}